# Dzyaloshinskii-Moriya Gradients Unlock Topological Dimensional Reduction in Magnetic Hopfions

**Kui Yuan[1], Hailong Shen[1], Jiawei Dong[1,2], Xin Zhang[1], Longqing Chen[3], Yihan Wang[3], Bo Liu[3], Qing Zhang[4], Zhengshang Wang[5], Wenbin Qiu[6], Xiaoyi Wang[1#], Bing Han[7#], Yang Qiu[2#], Kun Zhang[3], Xudong Cui[5]**

1 College of Electronie&Infermation, Southwest Minzu University, State Ethnic Affairs Commission, Chengdu 610041, China.

2 Pico center, SUSTech Core Research Facilities, Southern University of Science and Technology, Shenzhen 518055, China.

3 Key Laboratory of Radiation Physics and Technology of the Ministry of Education, Institute of Nuclear Science and Technology, Sichuan University, Chengdu 610064, china.

4 School of Physics, University of Electronic Science and Technology of China,Chengdu 611731, China

5 The School of Optoelectronic Science and Engineering,University of Electronic Science and Technology of China, Chengdu, 611731, China.

6 Department of Fundamental Courses, Wuxi Institute of Technology, WuXi 214121, China.

7 School of Materials Science and Engineering, Eastern Institute of Technology, Ningbo 315000, China

Correspendence: Xiaoyi Wang 80300024@swun.edu.cn,

Bing Han bhan@eitech.edu.cn,

Yang Qiu qiuy@sustech.edu.cn.

## Abstract

Three-dimensional magnetic solitons retain topological protection only while their spin field remains continuous. Here we show that chemical inhomogeneity can break this protection in a controlled way, converting a hopfion-like toroidal texture into an effectively two-dimensional skyrmion string. Tilt-dependent Lorentz transmission electron microscopy, electron energy-loss spectroscopy, and micromagnetic simulations of pristine, uniformly Y-doped, and nonuniformly Y-doped disordered $TiO_2$ nanoparticles embedded in a FeCrNiMn host reveal two regimes. Uniform Y doping enriches $Ti^{3+}$/oxygen-vacancy localization centers and stabilizes closed rings with Hopf invariant $\boldsymbol{Q}_{\mathrm{H}} \approx 1$. Nonuniform Y doping forms a $Ti^{4+}$-rich, vacancy-depleted boundary that creates a sharp $\boldsymbol{q}=\boldsymbol{D}/(2\boldsymbol{A})$ gradient and a weak-moment leakage channel. This coupled mismatch torque and continuity leakage split the ring, leaving a skyrmion-string remnant and a field-sensitive helicity texture.

## Introduction

Magnetic hopfions are three-dimensional topological solitons in which skyrmion strings close into toroidal loops. Unlike two-dimensional skyrmions, their topology is encoded by linking in the full spin field, providing internal degrees of freedom attractive for spin-orbitronic and topological-information schemes. Hopf-type solitons originate from early topological-soliton and knot-soliton theory [1,2]; magnetic hopfions have only recently been imaged in synthetic Pt/Co/Ir multilayers by soft-x-ray microscopy [3] and in B20-type FeGe by Lorentz microscopy [4]. A central unresolved question is not only how to stabilize these objects, but how they fail when disorder and chemical gradients interrupt the ideal spin field.

That failure is fundamentally topological. A two-dimensional skyrmion is described by an integer skyrmion charge $\boldsymbol{Q}$, whereas a three-dimensional hopfion is described by the Hopf invariant $\boldsymbol{Q}_H$, which measures the linking of preimages—equivalently, skyrmion-string lines [5,6]. Symmetry-breaking interfaces can reconstruct two-dimensional textures, as in meron-to-skyrmion transitions with fractional effective charge [7]. For a hopfion, however, opening the closed skyrmion-string loop removes the global map on which $\boldsymbol{Q}_H$ is defined. The opened state may then unwind to a trivial texture or retain a local skyrmion-string character. Distinguishing these limits is essential for understanding three-dimensional magnetic topology in chemically complex materials.

We investigate this breakdown in disordered $TiO_2$ nanoparticles embedded in a FeCrNiMn high-entropy-alloy matrix. Three regions are compared, pristine disordered $TiO_2$, uniformly Y-doped $TiO_2$, and nonuniformly Y-doped $TiO_2$. This comparison separates structural disorder, average dopant concentration, and dopant gradients. The Fe-based matrix provides a magnetic host that couples to defect-related oxide moments through stray-field, dipolar, and interfacial interactions. Uniform Y doping modifies the $Ti^{3+}$/oxygen-vacancy population and is consistent with an increased effective chiral parameter $\boldsymbol{q}=\boldsymbol{D}/(2\boldsymbol{A})$, where $\boldsymbol{D}$ is the effective Dzyaloshinskii–Moriya interaction and $\boldsymbol{A}$ is the exchange stiffness. Nonuniform Y doping instead creates a $Ti^{4+}$-rich, vacancy-depleted boundary. This boundary serves two coupled functions: the $\boldsymbol{q}$ gradient drives an interfacial mismatch torque, and the weak-moment boundary permits local spin-continuity leakage. Together they open the parent toroidal texture and split it into a field-sensitive helicity-rich branch and a robust skyrmion-string-like branch.

To rationalize this bifurcation, we treat the toroidal texture as a closed chiral magnetic tube with core radius a, ring radius L, and helical twist q = D/(2A) [4,8,9]:

$$L(\mathbf{r}) = \frac{\eta}{q(\mathbf{r})} = \frac{2\eta A(\mathbf{r})}{D(\mathbf{r})} \tag{1}$$

where η is a geometric factor of order unity that depends weakly on anisotropy and applied field. Equation (1) captures the central competition: increasing D strengthens the twist and contracts the torus, whereas increasing A stiffens the texture and expands it. Within a finite disorder window, local inversion-symmetry breaking, irradiation-induced interface disorder, and electronic localization can modify the effective DMI and exchange parameters [10–15]. At a boundary between high- and low-doping regions, we describe neighboring toroidal textures with different q values phenomenologically by an interfacial mismatch energy with effective stiffness $\Gamma_{ij} > 0$; this form is motivated by DMI-controlled chiral-domain-wall dynamics and DMI-induced finite-boundary effects [16,17]:

$$\begin{cases} E_{\text{int-high}} = \dfrac{\Gamma_{\text{int-high}}}{2} \displaystyle\int_{\Sigma_{\text{int-high}}} \left[q_{int}(\mathbf{s}) - q_{high}(\mathbf{s})\right]^2 \, dS \\ E_{\text{int-low}} = \dfrac{\Gamma_{\text{int-low}}}{2} \displaystyle\int_{\Sigma_{\text{int-low}}} \left[q_{int}(\mathbf{s}) - q_{low}(\mathbf{s})\right]^2 \, dS \\ E_{\text{high-low}} = \dfrac{\Gamma_{\text{high-low}}}{2} \displaystyle\int_{\Sigma_{\text{high-low}}} \left[q_{high}(\mathbf{s}) - q_{low}(\mathbf{s})\right]^2 \, dS \end{cases} \tag{2}$$

A larger $\boldsymbol{q}$ mismatch produces a larger interfacial torque; spin-continuity breakdown should therefore nucleate where the mismatch energy is largest, particularly adjacent to the high-DMI domain when $\boldsymbol{E}_{\text{int,high}} > \boldsymbol{E}_{\text{int,low}} > \boldsymbol{E}_{\text{high-low}}$. Each branch then carries both mismatch energy and elastic curvature energy.

For a slender chiral tube with $\boldsymbol{a}/\boldsymbol{L} < 1$, the exchange-dominated curvature cost scales as $\boldsymbol{W} \propto (\partial \boldsymbol{m})^2 \propto 1/\boldsymbol{L}^2$, where m is the local magnetization in tube coordinates [18]. Smaller toroidal textures therefore store more curvature energy. Under a sharp DMI gradient, the high-DMI side accumulates the larger excess energy, $\boldsymbol{E}_{\text{int,high}} + \boldsymbol{W}_{\text{high}} > \boldsymbol{E}_{\text{int,low}} + \boldsymbol{W}_{\text{low}}$, and is the first to unwind. The low-DMI side, with smaller mismatch and curvature costs, can remain below the threshold for complete unwinding and survive as a skyrmion-string-like remnant. The model thus predicts DMI-dependent ring contraction in chemically uniform regions and asymmetric ring opening across a sharp q gradient.

The micromagnetic simulation in Fig. 1(e) summarizes the proposed mechanism: a weak-moment interface acts as a leakage channel for local topological breakdown, allowing a linked toroidal texture to evolve into a skyrmion-string-like branch and a helicity-rich topologically trivial branch.

Magnetic textures were imaged in Fresnel-mode Lorentz transmission electron microscopy (LTEM) using a Talos 200X microscope operated at 200 kV under representative magnetic fields from −200 to +200 mT. Tilt series were acquired from 0°to 40° magnetic-domain maps were reconstructed from ±300 μm defocus stacks using PyLorentz transport-of-intensity-equation analysis [19], with drift correction and noise suppression. Micromagnetic simulations were performed with MuMax3 [20]. Pristine and uniformly Y-doped cases were modeled as effective defect-induced magnetic media embedded in a FeCrNiMn host, analytic Hopf seeds were relaxed, and only the FULL_FOR_$Q_H$ magnetization states were used for Hopf-charge evaluation. The nonuniformly doped case contained low-DMI and high-DMI branches separated by a $Ti^{4+}$-rich, oxygen-vacancy-depleted interface, where the $\boldsymbol{q}=\boldsymbol{D}/(2\boldsymbol{A})$ mismatch supplied the interfacial drive. These parameters are effective micromagnetic parameters for defect-induced textures rather than bulk constants of stoichiometric $TiO_2$, the simulations test morphology, stability, and trends rather than extract ab initio material constants.

## Results and discussion

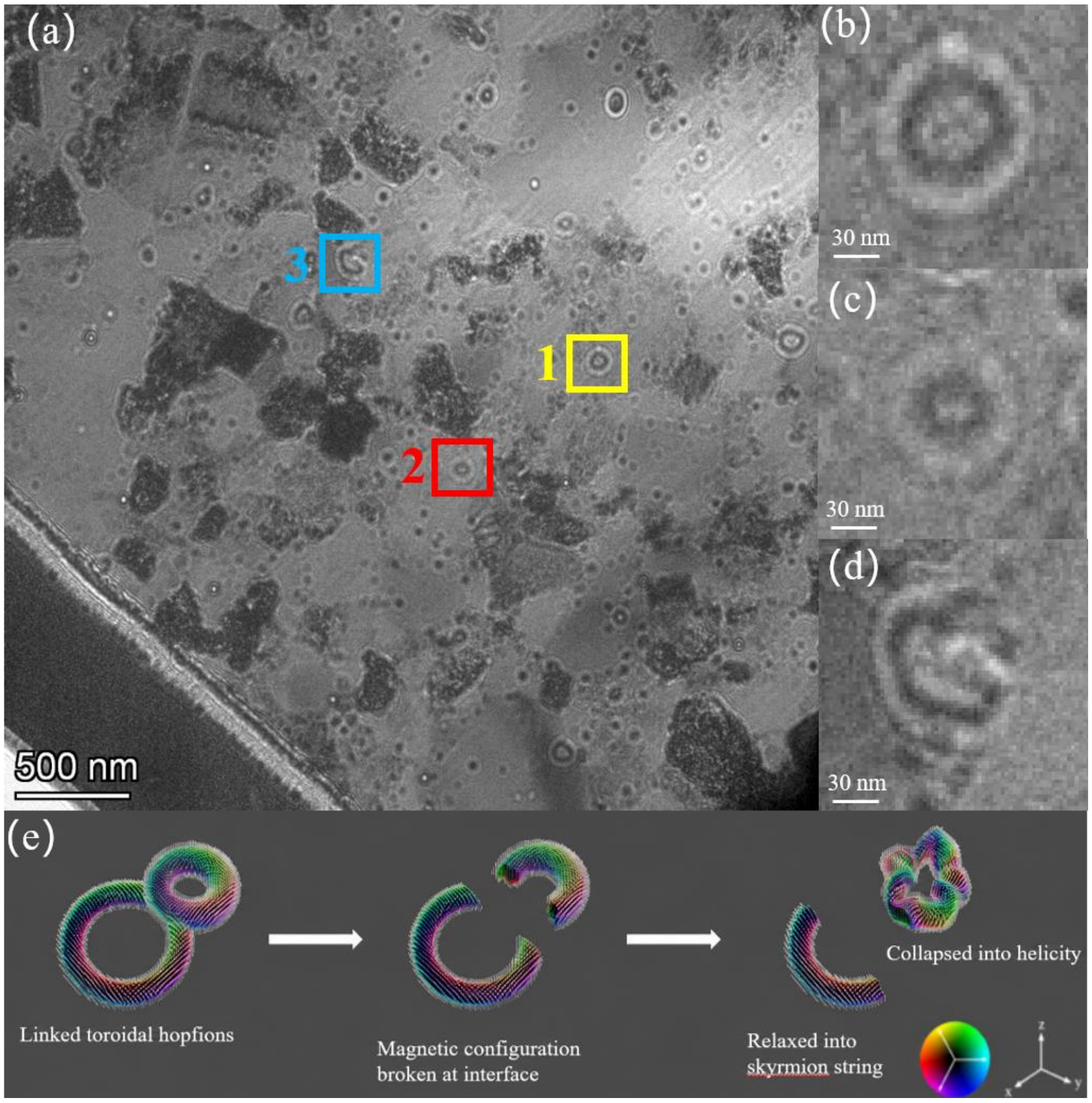

FIG. 1. Overview and proposed mechanism for DMI-gradient-assisted hopfion-ring breakdown. (a) Fresnel-mode LTEM overview of $TiO_2$ nanoparticles embedded in a FeCrNiMn high-entropy-alloy matrix, showing the three investigated regions: pristine $TiO_2$ (region 1), uniformly Y-doped $TiO_2$ (region 2), and nonuniformly Y-doped $TiO_2$

(region 3). (b)-(d) Magnified LTEM images of the pristine, uniformly Y-doped, and nonuniformly Y-doped regions, respectively. (e) Schematic mechanism. A linked toroidal hopfion-like configuration opens at a weak-moment interface and relaxes into a skyrmion-string-like branch and a helicity-rich topologically trivial branch.

Figure 1(a) shows the FeCrNiMn matrix containing $TiO_2$ nanoparticles. Three representative regions isolate the roles of disorder, uniform Y incorporation, and spatially nonuniform Y distribution. The magnified LTEM images in Figs. 1(b)–1(d) reveal the key phenomenology, pristine and uniformly Y-doped particles show ring-like magnetic contrast, whereas the nonuniformly Y-doped region displays an opened, discontinuous texture. This contrast indicates that a chemically uniform defect landscape supports a closed toroidal spin texture, whereas a dopant-gradient boundary interrupts three-dimensional spin continuity. Supporting structural and compositional evidence is provided in the Supplementary Materials (HRTEM, FFT, and elemental maps in Figs. S1–S2).

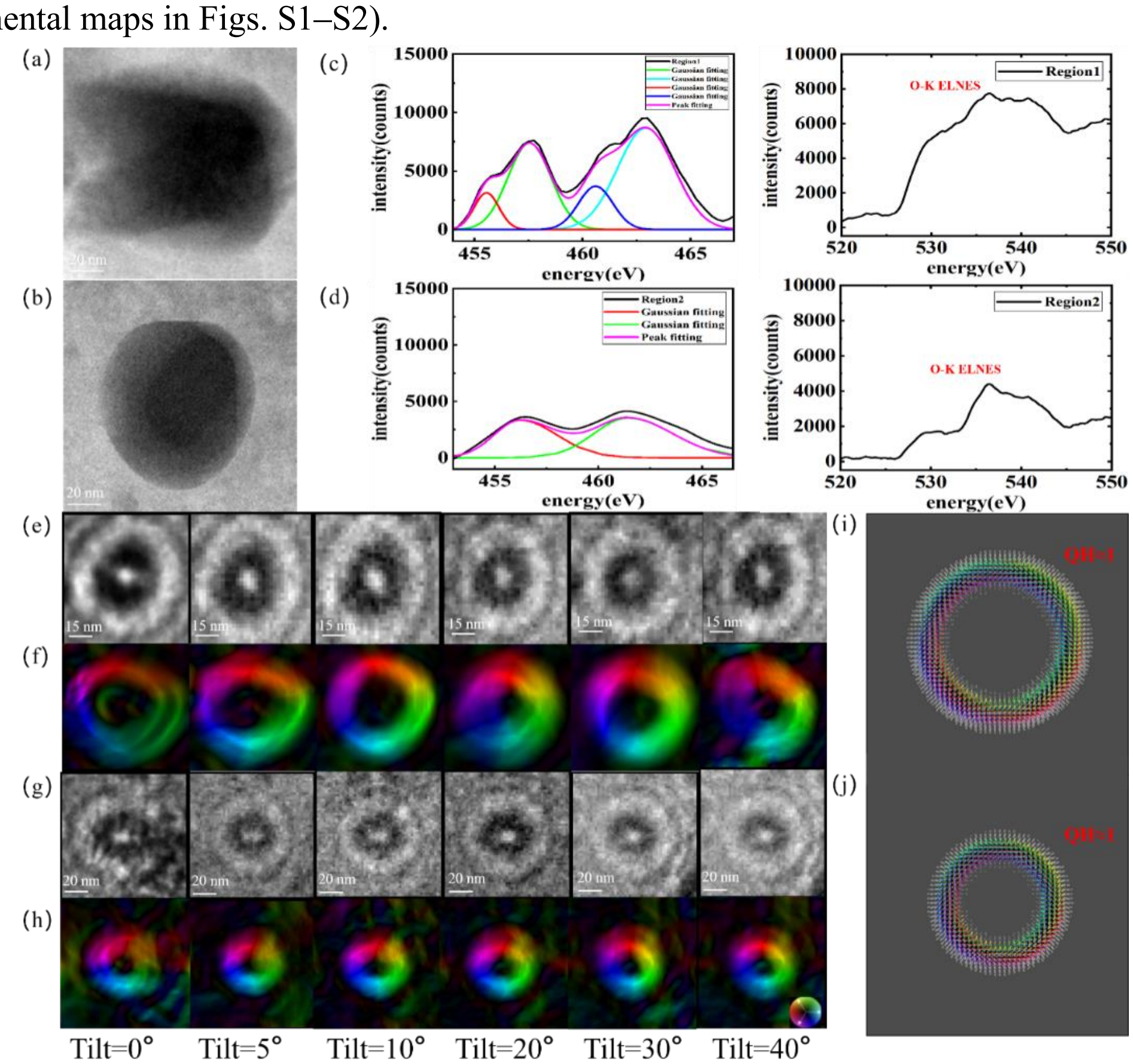

FIG. 2. Electronic structure and simulation-assisted assignment of closed toroidal textures in pristine and uniformly Y-doped $TiO_2$. (a,b) LTEM images of pristine and uniformly Y-doped $TiO_2$ regions. (c,d) Ti $L_{2,3}$-edge spectra (left) and O K-edge spectra (right), with corresponding fits, acquired from the pristine and uniformly Y-doped regions. (e,f) Tilt-dependent LTEM images and reconstructed magnetic-domain maps of pristine $TiO_2$. (g,h) Tilt-dependent LTEM images and reconstructed magnetic-domain maps of uniformly Y-doped $TiO_2$. (i,j) Effective MuMax3-simulated closed-ring states with QH close to 1 in pristine and uniformly Y-doped $TiO_2$, respectively.

Tilt-dependent LTEM/TIE reconstructions support the assignment of closed toroidal spin textures in the pristine and uniformly Y-doped regions. LTEM/TIE measures projection-sensitive magnetic contrast and does not uniquely determine $\boldsymbol{Q}_{\mathrm{H}}$, the weak contrast change across the measured tilt range [Figs. 2(e)–2(h)] provides morphological evidence for hopfion-like toroidal textures, while near-unity $\boldsymbol{Q}_{\mathrm{H}}$ values are obtained from the corresponding effective micromagnetic simulations. The representative ring diameter decreases from approximately 70 nm in pristine disordered $TiO_2$ to approximately 40 nm in uniformly Y-doped $TiO_2$, following the inverse $\boldsymbol{q}$ tendency predicted by Eq. (1). This contraction is consistent with increased effective $\boldsymbol{q}=\boldsymbol{D}/(2\boldsymbol{A})$ after Y incorporation and defect reconstruction. EELS identifies the associated electronic change. The O K-edge ELNES indicates oxygen vacancies in both pristine and uniformly Y-doped disordered $TiO_2$ [21,22], the Ti $L_{2,3}$ edge shows that the pristine region is dominated by $Ti^{4+}$ ($3d^0$), whereas the uniformly doped region contains an enhanced $Ti^{3+}$ ($3d^1$) fraction [21]. The unpaired d electron can localize in the disordered oxide and contribute to an effective magnetic moment. We therefore attribute the magnetic response to $Ti^{3+}$/oxygen-vacancy-related localized moments coupled to interfacial inversion-symmetry breaking and the FeCrNiMn host, rather than to intrinsic ferromagnetism of stoichiometric $TiO_2$. The higher density of $Ti^{3+}$/vacancy-related localization centers in uniformly doped $TiO_2$ can stiffen the effective local moment and increase the barrier against spin-winding disruption. The simulated closed-ring morphologies in Figs. 2(i) and 2(j), together with near-unity $\boldsymbol{Q}_{\mathrm{H}}$, support this defect and interface mediated effective-DMI stabilization scenario.

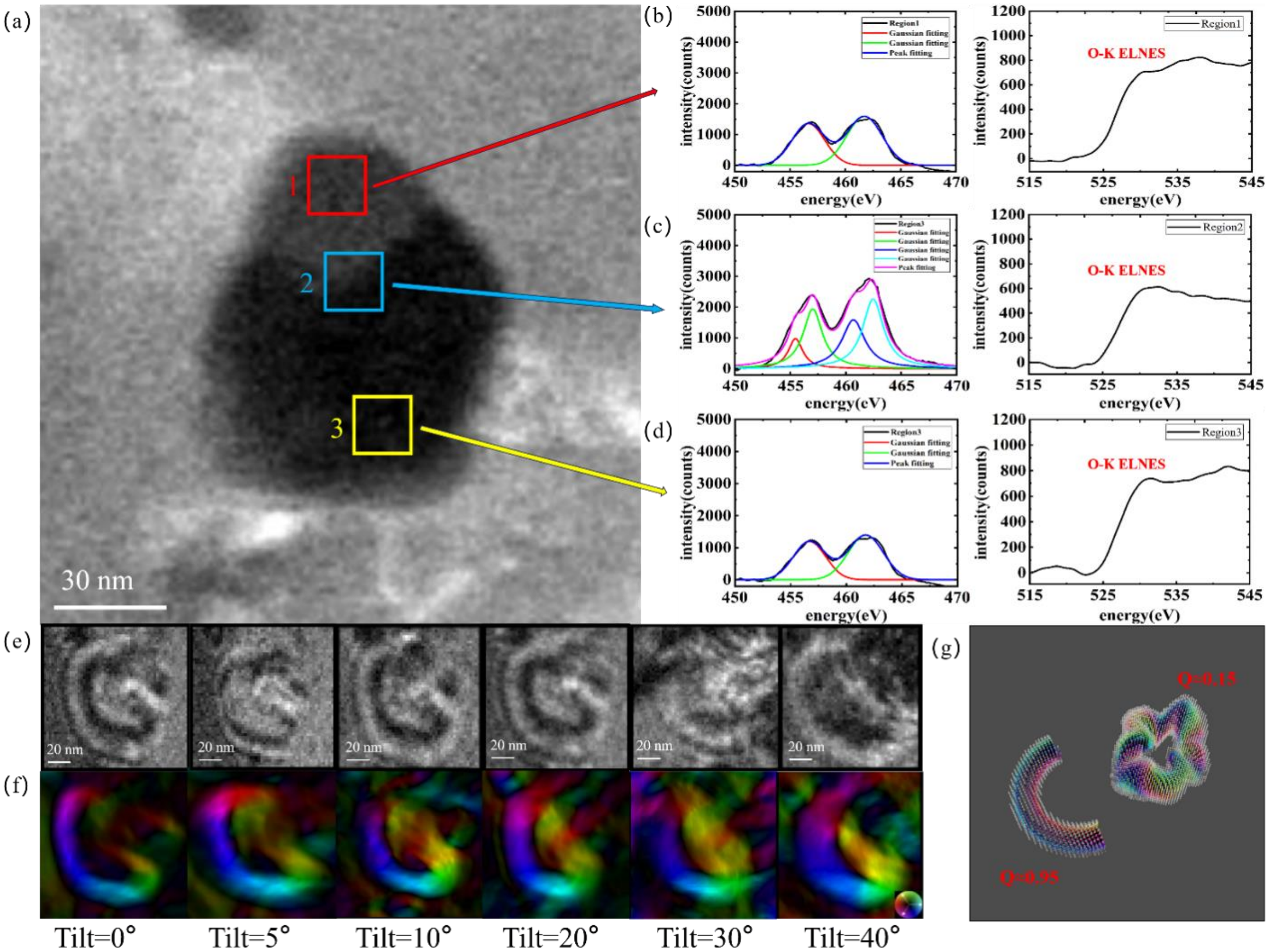

FIG. 3. Electronic and magnetic structure of a nonuniformly Y-doped $TiO_2$ nanoparticle. (a) HRTEM image showing the upper domain (region 1), interfacial boundary (region 2), and lower domain (region 3). (b)-(d) Ti $L_{2,3}$-edge spectra (left) and O K-edge spectra (right), with fits, acquired from the upper domain, boundary, and lower domain. (e,f) Tilt-dependent LTEM images and reconstructed magnetic-domain maps of the same nanoparticle. (g) MuMax3-simulated post-breakdown state, showing a low-DMI skyrmion-string-like remnant and a high-DMI helicity-rich topologically trivial branch.

Nonuniform Y doping shifts the role of disorder from stabilization to controlled failure (Fig. 3). EELS shows that both high- and low-Y domains remain enriched in $Ti^{3+}$ and contain oxygen-vacancy-related localization centers, whereas the boundary between them is $Ti^{4+}$ dominated and relatively vacancy depleted. The Ti $L_{2,3}$ fine structure at the boundary is closer to that of pristine $TiO_2$ than to either doped domain, indicating a local rebound toward $Ti^{4+}$. In the effective model, this relatively weak electron-localization boundary imposes a sharp q mismatch and generates an interfacial spin torque. The reduced $Ti^{3+}$/oxygen-vacancy-related moment density at this boundary implies weaker defect-induced local magnetic locking, consistent with the known role of $Ti^{3+}$ moments in oxygen-deficient $TiO_2$ [23]. We therefore model this boundary as a local spin-continuity leakage channel; such topology-changing events are consistent with defect- and boundary-mediated skyrmion annihilation pathways reported in three-dimensional and finite-size magnetic systems [24–26]. The q gradient is the driving field for deformation, whereas the weak-moment boundary is the gate through which the closed ring opens. The tilt-dependent magnetic-domain maps [Figs. 3(e)–3(f)] follow this picture, the closed toroidal texture seen in chemically uniform particles is absent on the high-Y side, whereas the low-Y side retains a partial loop with reduced curvature. To test this pathway, two coupled hopfion-like rings with different diameters were initialized in a nonuniform effective-DMI landscape containing a sharp q gradient and a $Ti^{4+}$-rich weak-moment boundary, as schematized in Fig. 1(e). Free relaxation produced the asymmetric post-breakdown state in Fig. 3(g). Because the opened texture is no longer a closed three-dimensional spin map, the global Hopf invariant is not defined after breakdown. We therefore calculated an effective cross-sectional skyrmion charge for each branch. The high-DMI branch has a small effective charge, $Q = 0.15$, consistent with a topologically trivial helicity-rich state. The low-DMI tube-like remnant retains $Q = 0.95$, showing that local skyrmion-string-like topology survives after the parent Hopf texture opens.

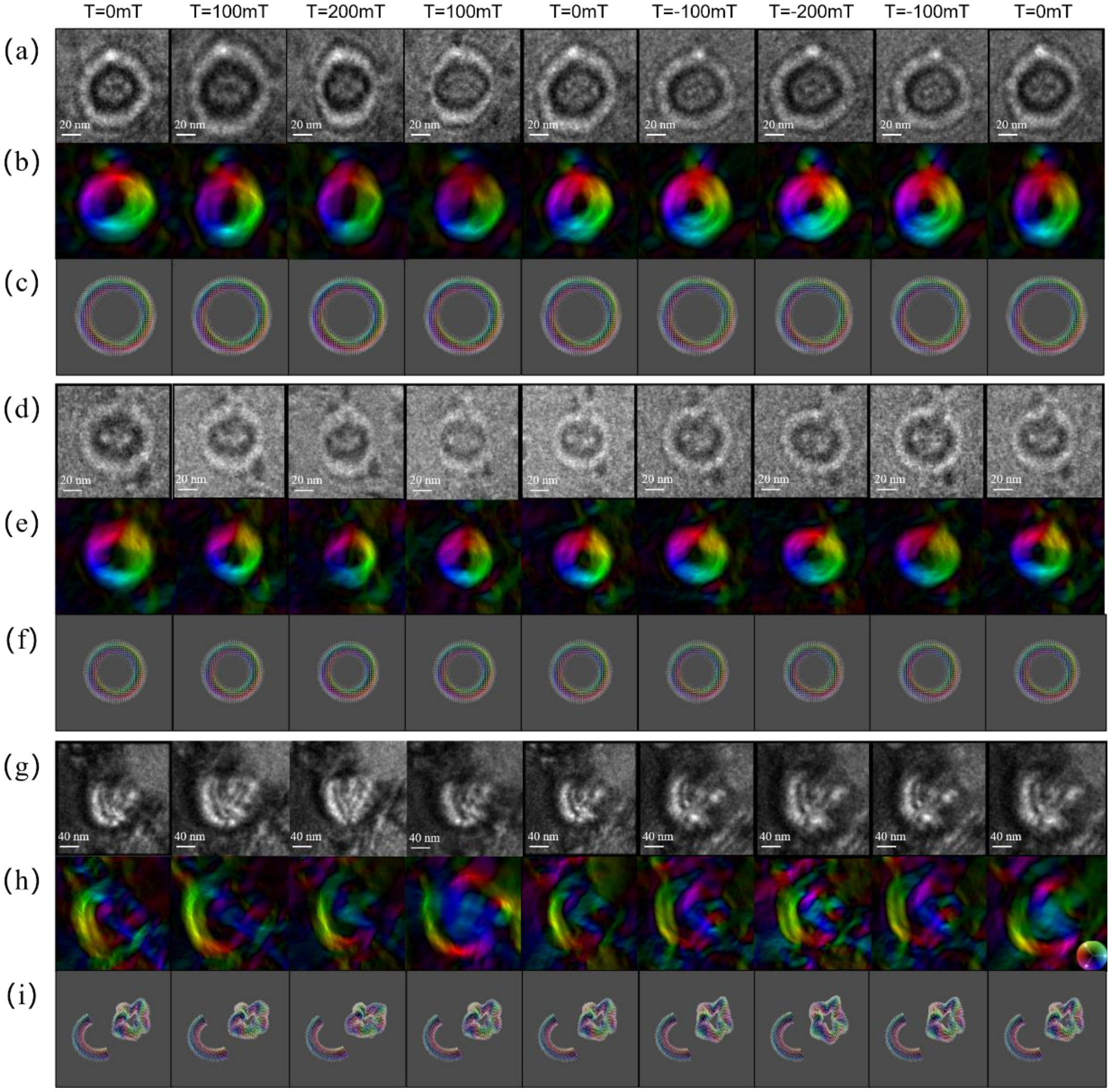

FIG. 4. Field response of closed and post-breakdown magnetic textures during a representative hysteresis-like field sweep at zero tilt. (a)-(c) LTEM images, reconstructed magnetic-domain maps, and MuMax3 simulations of the pristine $TiO_2$ region under varying fields. (d)-(f) Corresponding images, reconstructions, and simulations of the uniformly Y-doped region. (g)-(i) Corresponding images, reconstructions, and simulations of the nonuniformly Y-

doped region.

The field response separates robust toroidal textures from the two post-breakdown branches (Fig. 4). In the pristine and uniformly Y-doped regions [Figs. 4(a)–4(f)], the hopfion-like rings remain nearly unchanged throughout the representative field sweep, despite the stray fields of the Fe-based matrix and the applied magnetic excitation. The corresponding simulations reproduce this stability. Because finite mesh size, open boundaries, and dissipative Landau-Lifshitz-Gilbert dynamics in MuMax3 do not strictly conserve topological invariants [20], we introduced boundary compactification, DMI-induced chiral locking, and exchange-anisotropy confinement barriers to suppress artificial leakage in the initial closed rings; these choices are motivated by reported finite-size, boundary, and multiscale pathways for skyrmion stability and topological transformation [17,25,26]. The simulated Hopf invariant remains close to unity during the field cycle, consistent with robust closed-ring stability and with the small experimental changes in ring morphology [Figs. 4(c) and 4(f)]. The nonuniformly doped particle behaves differently. The high-Y branch [Figs. 4(g)–4(i)] is erratic and follows the reversal of the surrounding FeCrNiMn matrix, as expected for a topologically trivial, field-sensitive high-DMI branch. The low-Y branch remains string-like and stable throughout the same field cycle, showing that two-dimensional skyrmion-string-like robustness survives after leakage-assisted opening of the parent torus. Simulations initialized from the freely relaxed post-breakdown state reproduce this bifurcation: a low-DMI remnant with effective Q in the range 0.9–1.0 coexists with a high-DMI helicity-rich texture whose effective charge remains below approximately 0.2. The field response therefore provides the experimental signature of the two energy pathways predicted by Eqs. (1) and (2).

## Conclusion

We show that a chemical gradient can reduce a closed three-dimensional hopfion-like ring to lower-dimensional magnetic textures when coupled to a weak-moment interfacial leakage channel. Uniform Y doping in disordered $TiO_2$ inclusions increases $Ti^{3+}$/oxygen-vacancy-related localization centers and supports compact, field-robust toroidal textures with simulated $\boldsymbol{Q}_{\mathrm{H}}$ close to 1. Nonuniform Y doping instead creates a $Ti^{4+}$-rich, vacancy-depleted boundary that both generates a sharp $\boldsymbol{q}=\boldsymbol{D}/(2\boldsymbol{A})$ mismatch and weakens local chiral locking. The toroidal spin field then loses continuity and opens asymmetrically, the high-DMI side relaxes into a field-sensitive helicity rich trivial state, whereas the low-DMI side retains a skyrmion string like remnant with effective charge near unity. These results show that chemical gradients and weak-moment interfaces can reshape rather than simply destroy three dimensional magnetic topology, providing a route to controlled transition of magnetic topological dimensionality in defect-rich spin-orbitronic materials.

Acknowledgments: This work was supported by the National Natural Science Foundation of China (No. 12305315, 52573248), the National Key Research and Development Program of China(No.2025YFH0102004) and the Sichuan Provincial Science and Technology Program (No. 2026YFHZ0153, 2025ZDZX0021).

## Supplemental material

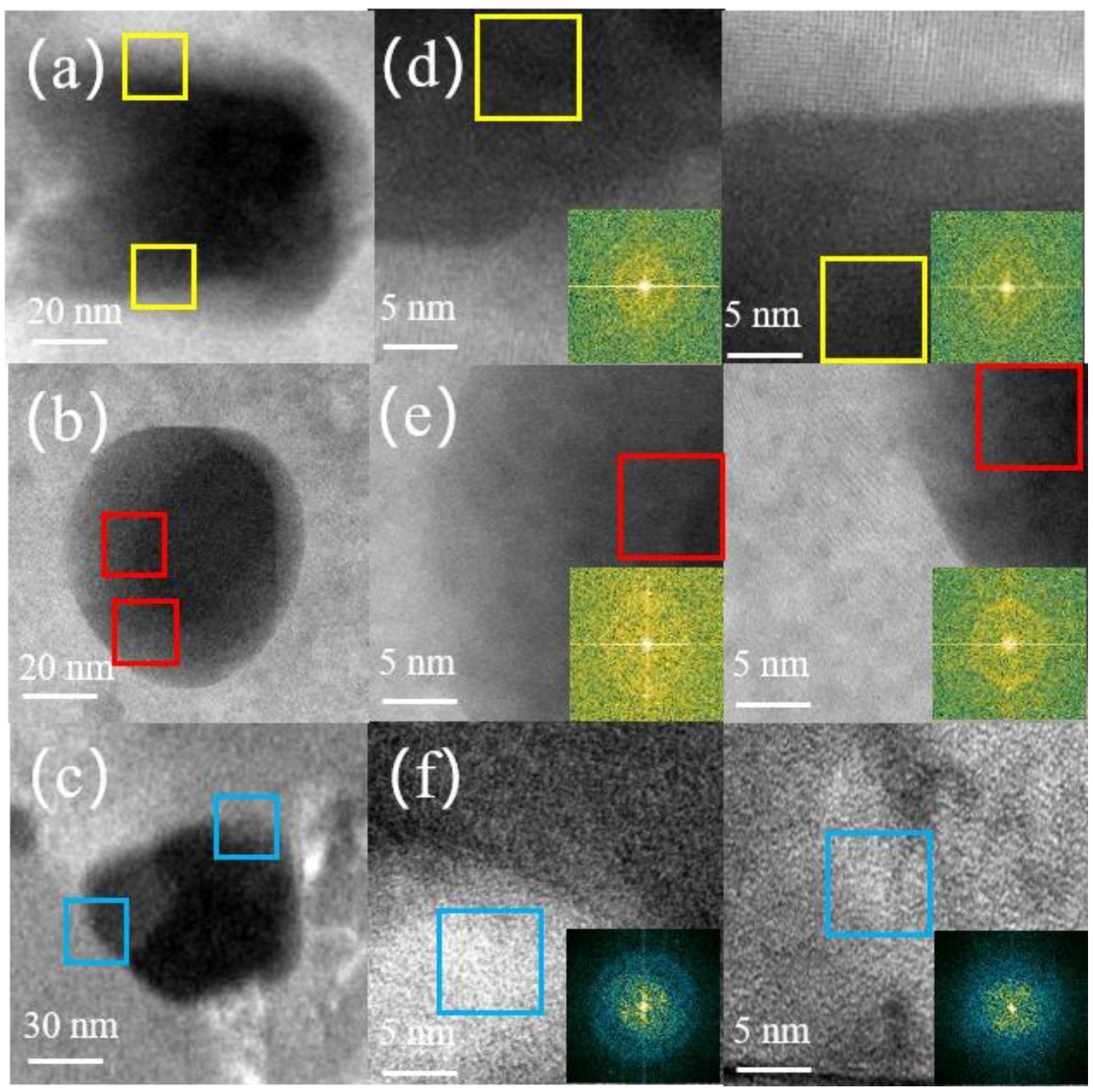

FIG. S1. Local structural characterization of pristine, uniformly Y-doped, and nonuniformly Y-doped $TiO_2$ nanoparticles. (a)-(c) HRTEM images showing structurally disordered $TiO_2$ nanoparticles embedded in the FeCrNiMn matrix. (d)-(f) Enlarged HRTEM views of the marked regions with corresponding FFT patterns, confirming local structural disorder and nanoscale lattice variation.

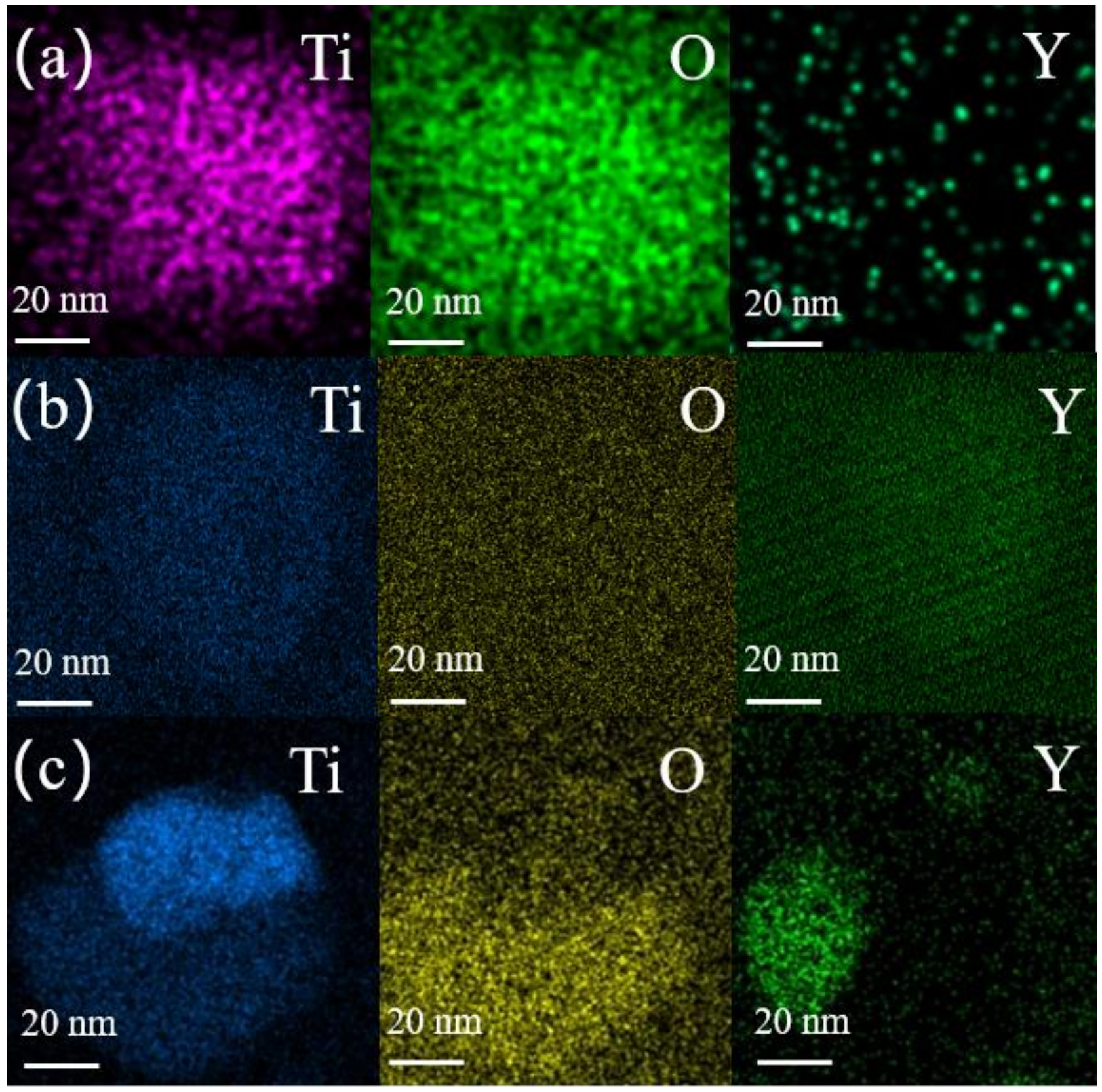

FIG. S2. Elemental distribution of Ti, O, and Y in pristine, uniformly Y-doped, and nonuniformly Y-doped $TiO_2$ regions. (a) Ti, O, and Y maps of the pristine $TiO_2$ region. (b) Ti, O, and Y maps of the uniformly Y-doped $TiO_2$ region. (c) Ti, O, and Y maps of the nonuniformly Y-doped $TiO_2$ region, showing spatially nonuniform Y distribution.

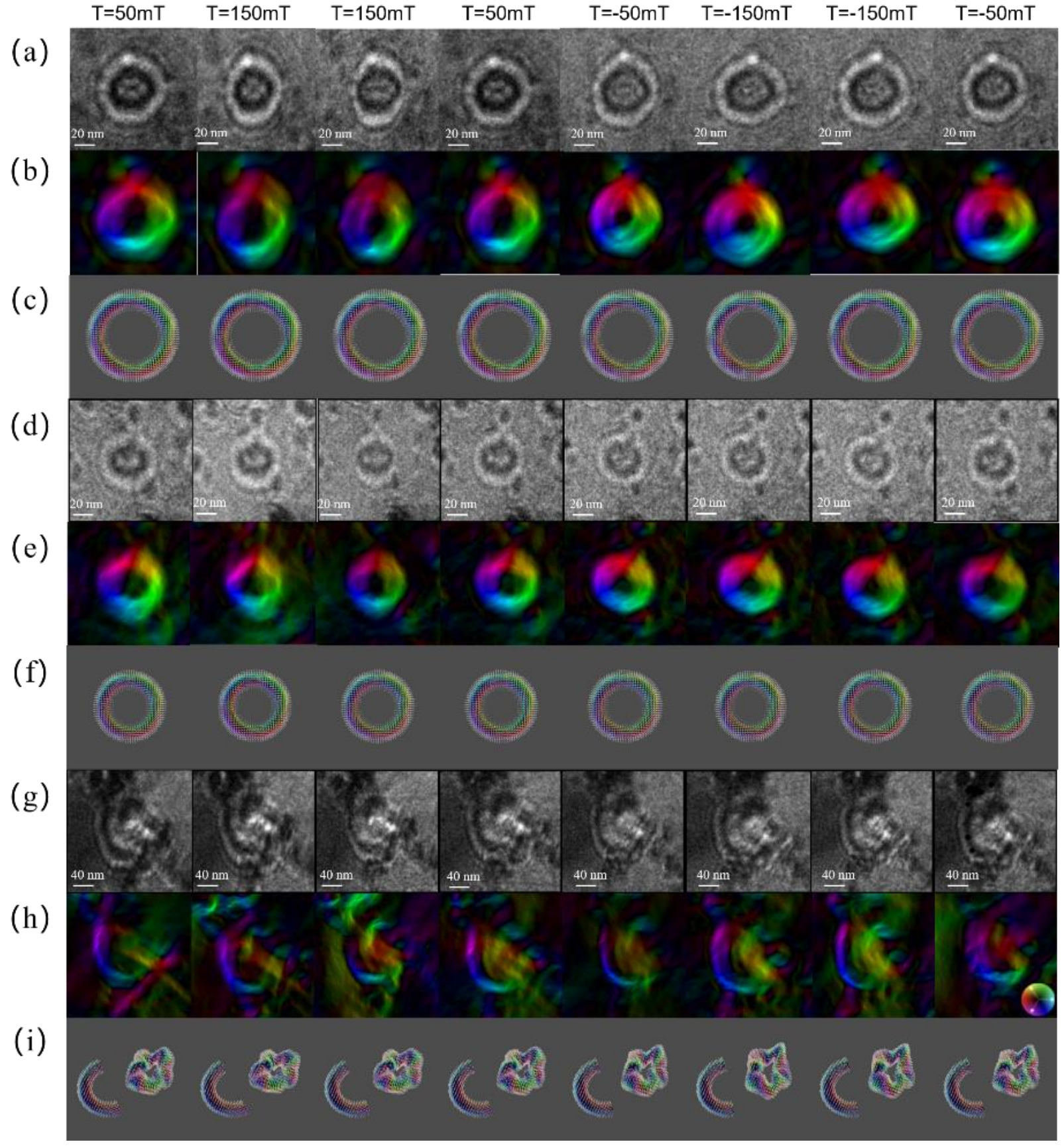

FIG. S3. (a)-(c) LTEM images, reconstructed magnetic-domain maps, and MuMax3 simulations of the pristine $TiO_2$ region under varying fields. (d)-(f) Corresponding images, reconstructions, and simulations of the uniformly Y-doped region. (g)-(i) Corresponding images, reconstructions, and simulations of the nonuniformly Y-doped region.